\documentclass[iop]{emulateapj} 
\usepackage{graphicx}

\def\edt#1{#1}

\def\HI{\ion{H}{1}}

\def\eh2{\ensuremath{e_{\mathrm{H2}}}}
\def\nh2{\ensuremath{n_{\mathrm{H2}}}}

\def\h2{\ensuremath{\mathrm{H}_2}}

\def\HI{\ion{H}{1}}
\def\HeI{\ion{He}{1}}
\def\HeII{\ion{He}{2}}
\def\HeIII{\ion{He}{3}}
\def\CaII{\ion{Ca}{2}}

\def\Bifrost{{\it Bifrost}}
\def\radyn{{\it RADYN}}
\def\multitd{{\it Multi3d}}
\def\multi{{\it MULTI}}
\def\figspath{.}
\def\Halpha{\mbox{H\hspace{0.1ex}$\alpha$}}
\def\NaD{\ion{Na}{1}\,D} 
\def\rmit#1{{\it #1}}        
\def\eg{\rmit{e.g.,}}  
\def\rma{{\rm a}} 
\def\rmb{{\rm b}} 
 
\def\rmd{{\rm d}}  
\def\rme{{\rm e}}

 \def\rmH{{\rm H}}

\def\rmt{{\rm t}}

\def\cf{cf.}
\def\ie{\rmit{i.e.,}}   

\begin{document}

\title{The formation of the \Halpha\ line in the solar
chromosphere}
  
   \author{J.~Leenaarts$^{1,2}$}\email{jorritl@astro.uio.no}
   \author{M.Carlsson$^{1,3}$}\email{mats.carlsson@astro.uio.no}
   \author{L. Rouppe van der Voort$^1$}\email{rouppe@astro.uio.no} 

\affil{$^1$ Institute of
  Theoretical Astrophysics, University of Oslo, P.O. Box 1029
  Blindern, N--0315 Oslo, Norway}
\affil{$^2$ Sterrekundig Instituut, Utrecht University, P.O. Box 80\,000
         NL--3508 TA Utrecht, The Netherlands}
\affil{$^3$ Center of Mathematics for Applications,
  University of Oslo, P.O. Box 1053
  Blindern, N--0316 Oslo, Norway}

   \date{Received; accepted}

   \begin{abstract}
 
We use state-of-the-art radiation-MHD simulations and 3D non-LTE
radiative transfer computations to investigate \Halpha\ line formation
in the solar chromosphere and apply the results of this investigation
to develop the potential of \Halpha\ as diagnostic of the chromosphere.

We show that one can accurately model \Halpha\ line formation assuming
statistical equilibrium and complete frequency redistribution provided
the computation of the model atmosphere included non-equilibrium
ionization of hydrogen, and the Lyman-$\alpha$ and Lyman-$\beta$ line
profiles are described by Doppler profiles. 

We find that 3D radiative transfer is essential in modeling hydrogen
lines due to the low photon destruction probability in \Halpha. The
\Halpha\ opacity in the upper chromosphere is mainly sensitive to the
mass density and only weakly sensitive to temperature. 

We find that the \Halpha\ line-core intensity is correlated with the
average formation height: \edt{ the larger the
average formation height, the lower the intensity}. The line-core width is a measure of the gas temperature in the
line-forming region. The fibril-like dark structures seen in
\Halpha\ line-core images computed from our model atmosphere are
tracing magnetic field lines. These structures are caused by
field-aligned ridges of enhanced chromospheric mass density that raise
their average formation height, and therefore makes them appear dark
against their deeper-formed surroundings.
We compare with
observations, and find that the simulated line-core widths are very
similar to the observed ones, without the need for additional microturbulence.

\end{abstract}

  
\section{Introduction}                          \label{sec:introduction}

The \Halpha\ line is one of the most popular lines for studying the
solar chromosphere. In fact, the chromosphere was defined as what is
seen in this line
\citep{1868RSPS...17..128L}. 
Yet, nearly 150 years after its discovery there is still much
unknown about the chromosphere and the formation of \Halpha. For an overview of the current
state of understanding and open questions of \Halpha\ as a diagnostic
of the chromosphere we refer to an excellent series of discussions by
\citet{2008ASPC..397...54R,
  2010rast.conf..163R,2010MmSAI..81..565R,2011arXiv1110.6606R}. 

The lack of understanding is due to the particularly complicated
physics of this layer. It forms the interface between the convection
zone and the corona and chromospheric modeling therefore has to
include at least the upper convection zone and the lower corona. The
presence of magnetic fields leads to structures that cannot be
reproduced in one or two-dimensional modeling. Three-dimensional
geometry is therefore required, which significantly increases the
amount of computational work required.

Over the last few years it has become possible to create models of the
chromosphere that include much of the required physics and the
required physical extent: they model the
atmosphere from the upper convection zone to the lower corona, and
include resistive magnetohydrodynamics, full 3D radiation in
the photosphere and lower chromosphere, parameterized radiative losses
in the upper chromosphere and corona, thermal conduction and a
realistic equation of state. 
\citep{2004IAUS..223..385H,2007ASPC..368..107H,2011A&A...531A.154G}
These models have been used to study
various chromospheric phenomena, such as
dynamic fibrils
\citep{2006ApJ...647L..73H,2007ApJ...655..624D},
the presence of Alfv\'enic waves
\citep{2007Sci...318.1574D},
magnetic flux emergence
\citep{2008ApJ...679..871M, 2009ApJ...702..129M}
and spicules
\citep{2009ApJ...701.1569M,2011ApJ...736....9M}

All these studies compared the time-evolution of the MHD variables in
the simulations with some observed quantity, rather than first
computing synthetic observations. For example, the dynamic fibril
studies compared observed motions of structures observed in
\Halpha\ to motions of a temperature iso-contour in the numerical
simulations.

Direct comparison of synthetic observables generated from simulations
with observations have been more scarce. This is partly caused by the
complex radiative transfer processes in the chromosphere. Spectral
lines form in non-LTE, partial redistribution of photons (PRD) is
important for many lines, and the effects of full 3D radiation require
complicated and computing-intensive radiative transfer
calculations. Examples of such computations are the work on bisectors
of the \NaD\ and \CaII\ 8542\,{\AA} lines by
\citet{2006ApJ...639..516U},
who used a 3D hydrodynamical model that extended up to 1000\,km
above optical depth unity, covering the lower and mid chromosphere
only.
\citet{2009ASPC..415...87L}
and
\citet{2009ApJ...694L.128L,2010ApJ...709.1362L}
investigated the formation of \NaD, \CaII\ 8542\,\AA\ and \CaII\,H in
3D MHD models that extended up into the corona. They found that 3D
effects are important in the cores of all lines. The 3D effects reduced the
overall contrast of line core images, and for \NaD\ it was found that
the bright halos around magnetic field concentrations are caused by
photon scattering.

The previous studies found reasonable agreement between observations
and synthetic imagery for the low-forming \NaD\ line. Agreement for
the \CaII\ 8542\,\AA\ was reasonable for quiet regions without
overlying fibrils.

As discussed by
\citet{2010MmSAI..81..576L}
the discrepancy between observed and synthetic \Halpha\ was big:
computations treating each column as a plane-parallel atmosphere
(henceforth 1D radiative transfer) showed a granulation pattern in the
line core. This is in stark contrast to the observations that show
either fibrils or the signature of acoustic shocks
\citep{2008SoPh..251..533R}.
It was speculated that this discrepancy could be caused by a lack of
spatial extent and spatial resolution of the employed simulations,
lack of treatment of non-equilibrium level populations (with non-equilibrium
it is meant that the time derivative of the level populations is not
necessarily equal to zero) effects and/or the
neglect of PRD.

In the current paper we report an investigation of the formation of
the \Halpha\ line. In Sec.~\ref{sec:tdtransfer}--\ref{sec:3Dtransfer} we show that in
order to accurately model \Halpha\ line formation one has to use a
model atmosphere computed with an equation of state (EOS) that takes  into account the
effect of non-equilibrium hydrogen ionization and one has
to compute the radiation field in full 3D. We also show that one can
neglect the time-dependence of the rate equations and use statistical
equilibrium in the subsequent non-LTE spectrum synthesis 
and that one can approximate PRD effects by treating the
Lyman lines with Doppler profiles in complete redistribution. In
Sec.~\ref{sec:analysis}--\ref{sec:discussion} we analyze the results
of our 3D computation in detail, compare with observations and finish
with a discussion of the results.

\section{The effect of non-equilibrium transfer on \Halpha} \label{sec:tdtransfer}

\begin{figure}
  \includegraphics[width=\columnwidth]{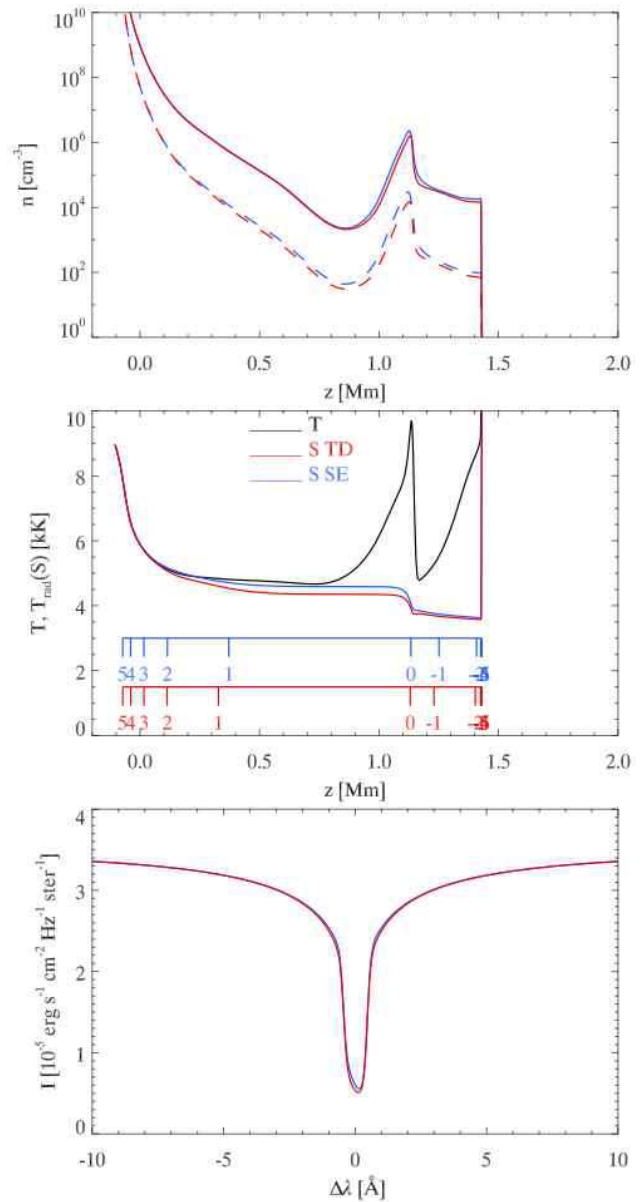}
  \caption{Comparison of non-equilibrium radiative transfer (red
    curves) to radiative transfer assuming statistical equilibrium
    (blue curves) in a one-dimensional model atmosphere computed with
    {\it Radyn}. Top: level population densities of the lower (solid)
    and upper (dashed) level of the \Halpha\ line as function of
    height. Middle: gas temperature (black) and the \Halpha\ line
    source function as function of height. The additional blue and red
    scales show $\log_{10} \tau$ at line center. Bottom: Vertically
    emergent \Halpha\ line profile.}
  \label{fig:radynvsz}
\end{figure}

\begin{figure}
  \includegraphics[width=\columnwidth]{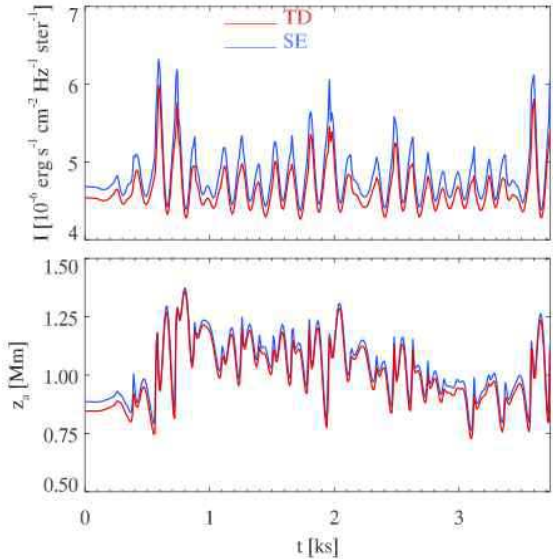}
  \caption{Comparison of non-equilibrium radiative transfer (red curves) to
    radiative transfer assuming statistical equilibrium (blue curves)
    in a time-series of
    one-dimensional model atmospheres computed with {\it Radyn}. Top:
    vertically emergent intensity in the \Halpha\ line
    core. Bottom: average formation height.
  \label{fig:radynvst}}
\end{figure}

To investigate the effect of non-equilibrium radiative transfer on the
formation of the \Halpha\ line we use a time-series of hydrodynamical
snapshots computed with the \radyn\ code
\citep{1992ApJ...397L..59C,
  1995ApJ...440L..29C,
  1997ApJ...481..500C,
2002ApJ...572..626C}.

This code solves the equations of conservation of mass, momentum,
energy and mass. In addition it solves the non-LTE radiative transfer
problem including non-equilibrium rate equations for hydrogen, helium
and calcium. It uses a 5-level plus continuum model atom for \HI\ and
\CaII, and a nine-level helium atom with 5, 3 and 1 levels in \HeI,
\HeII\ and \HeIII\ respectively. These equations are solved implicitly
on an adaptive mesh that moves points to regions where high spatial
resolution is required
\citep{1987JCoPh..69..175D}. 
\radyn\ solves the non-equilibrium radiative transfer problem for
hydrogen self-consistently including its effect on the ionization
balance and therefore on the EOS. It treats PRD
effects in the Lyman lines rather crudely, by truncating the Lyman
line profiles at 48\,km\,s$^{-1}$ away from the line-center
frequency. We show in Sec.~\ref{sec:prdtransfer} that this simple method
has only a minor effect on the formation of \Halpha.

In our \radyn\ simulation the lower boundary is located just
below the photosphere at a column mass of $8.1$\,g\,cm$^{-2}$,
roughly corresponding to $\tau_{500\mathrm{nm}}=16$. The upper
boundary is located in the corona, at 10,000~km above the lower
boundary. The bottom boundary is closed and driven with a prescribed
velocity field derived from MDI observations of the Doppler shifts of
the 676.78 nm \ion{Ni}{1} line in a quiet sun region. This driver
excites waves that travel upward and steepen into shocks. We ran the
simulation for slightly more than 1~hour of solar time and saved
output at 10\,s intervals, yielding a series of 374 snapshots of both
the hydrodynamical state and the pertinent radiative transfer
quantities.

For each of these 1D snapshots we then solved the non-LTE radiative
transfer problem for hydrogen assuming statistical equilibrium with
the radiative transfer code \multi\
\citep{1986UppOR..33.....C}
using identical hydrodynamic quantities (temperature, mass density,
electron density and vertical velocity) and model atom.
\multi\ and \radyn\ have identical source code for the routines
that compute the radiative transfer. We thus ensured that any
differences in the results are caused by the change from
non-equilibrium (NE) level populations to transfer assuming statistical
equilibrium (SE),
and not by differences in the numerical method.

Figure~\ref{fig:radynvsz} compares the formation of the \Halpha\ line
between NE and SE calculations at $t=3740$~s. At this time there is a
strong shock in the chromosphere, a circumstance which typically
exhibits the largest differences in line formation between NE and
SE. At times that the chromospheric temperature structure is smooth
there is almost no difference between the two assumptions.

The top panel shows the population densities of its upper and lower
level. In SE both populations are slightly higher than in the NE
case. Their ratio is however nearly equal, leading to a very similar
source function, as shown in the middle panel. The SE computation
shows a slightly higher source function throughout the simulated
chromosphere. The line-core optical depth scales are nearly equal. The
bottom panel compares the line profiles. They are near-identical,
except in the line core, where the SE intensity is larger, which
reflects the higher source function.  

In Fig.~\ref{fig:radynvst} we compare the time evolution of the
line-center intensity and the line-center average formation
height $z_\rma$. The latter is defined as
\begin{equation}
  z_\rma = \frac{\int_{z_\rmb}^{z_\rmt} z' \chi S \rme^{-\tau} \rmd z'}
  {\int_{z_\rmb}^{z_\rmt} \chi S \rme^{-\tau} \rmd z'},
\end{equation}
where $z_\rmb$ and $z_\rmt$ are the height of the bottom and top
boundary in the model and $\chi$, $S$ and $\tau$ are the line-center
opacity, source function and optical depth, respectively. The average
formation height is in general not a good measure of where the
\Halpha\ line-core intensity is formed: the contribution functions are
often multiply peaked and can be non-zero over ranges of more than
1\,Mm. However, if the formation height range has to be quantified by
one number, the average formation height is a better measure than the
height of optical depth unity.

The curves for SE closely follow the corresponding curves for the full
non-equilibrium solution. The SE intensity is consistently higher than
the NE intensity, varying between 10\% higher in the presence of shocks to
only 2\% higher during inter-shock phases. The SE average formation
height is typically 30~km higher during inter-shock phases, and 80~km
higher during shock passages. Both the differences in intensity and
average formation height are much smaller than the intrinsic
variability caused by the changing state of the atmosphere.

Therefore, we conclude that it is possible to assume SE for the
radiative transfer of hydrogen, provided the temperature and electron
density in the model atmosphere are computed with a simulation that
includes the effect of non-equilibrium hydrogen ionization. The latter
is essential: assuming statistical equilibrium when calculating the atmosphere leads to temperatures that can be more
than 2000\,K off and electron densities that can be even orders of
magnitude off
\citep{2002ApJ...572..626C,2007A&A...473..625L}.

This result is crucial in making it possible to compute realistic
synthetic imagery in hydrogen lines from 3D simulations of the solar
chromosphere. The detailed non-equilibrium treatment of radiative
transfer as in \radyn\ is too computationally demanding in
multi-dimensional geometry. It is, however, possible to compute the
non-equilibrium ionization balance of hydrogen and its effect on the
equation-of-state using approximations that are efficient enough to be
included in a 3D MHD code.

\section{3D simulation snapshot and 3D radiative transfer} 
\label{sec:bifrostmulti3d}

\begin{figure*}
  \includegraphics[width=\textwidth]{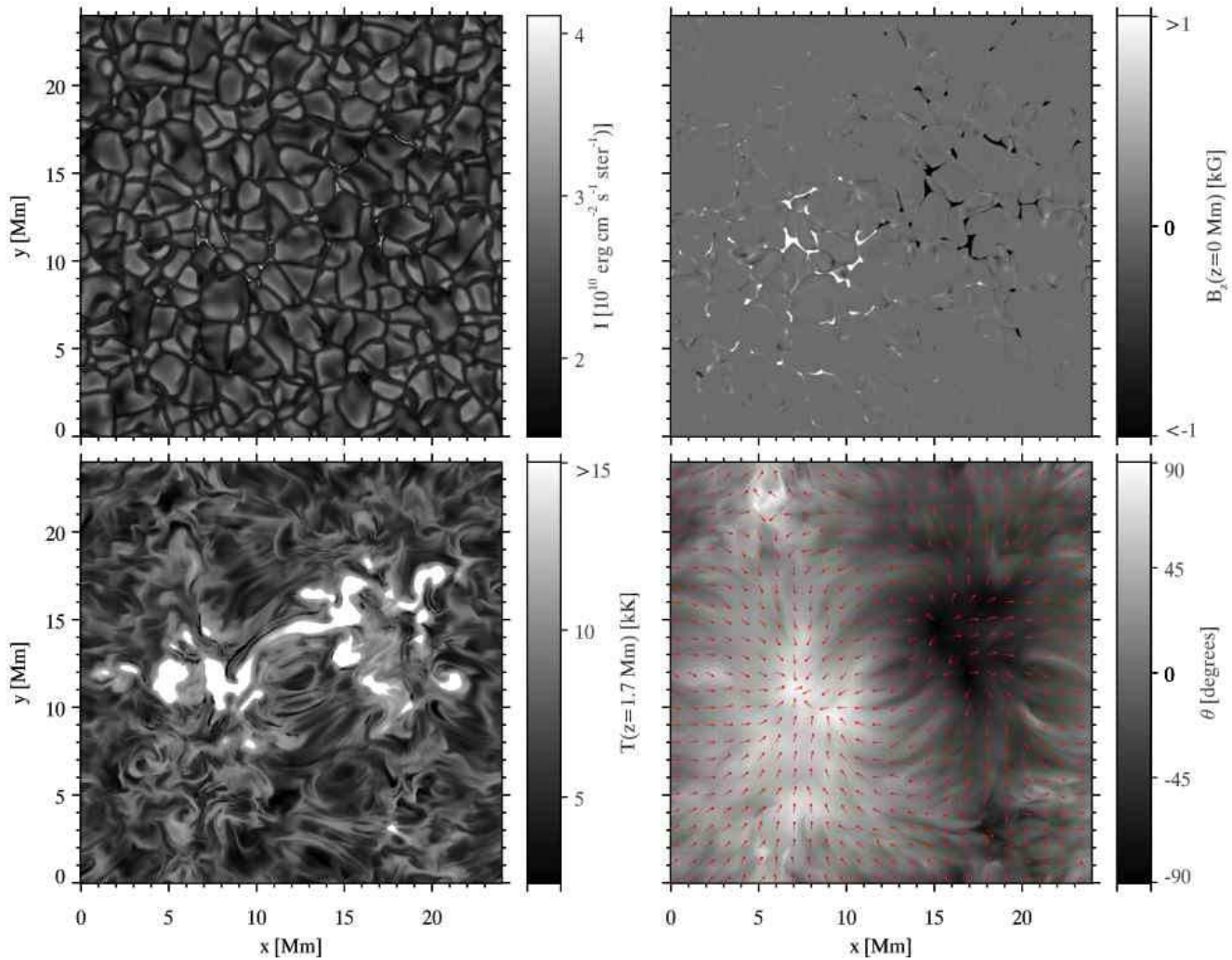}
  \caption{Characterization of the \Bifrost\ snapshot. Upper left:
    vertically emergent intensity in the continuum radiation bin.
    Upper right: Vertical magnetic field at $z=0$\,Mm. Lower left: Gas
    temperature at $z=1.7$\,Mm. Lower right: Angle of the magnetic
    field vector with the horizontal, with the red arrows indicating
    the direction of the horizontal component of the magnetic field. }
  \label{fig:snapcharac}
\end{figure*}

In Sec.~\ref{sec:prdtransfer} and~\ref{sec:3Dtransfer} we investigate
the influence of partial redistribution in Lyman lines and 3D
evaluation of the radiation field on the formation of \Halpha. We
study these effects in a snapshot of a 3D radiation-MHD simulation
computed with the \Bifrost\ code.

\Bifrost\ solves the equations of resistive MHD on a staggered
Cartesian grid. In addition to this the code can solve various
additional physical processes. The snapshot we are using was computed
with the following additional physics: Optically thick radiative
transfer in the photosphere and low chromosphere, parameterized
radiative losses in the upper chromosphere, transition region and
corona, thermal conduction along magnetic field lines and EOS that
includes the effects of non-equilibrium ionization of hydrogen.  As
discussed in Sec.~\ref{sec:tdtransfer}, the latter is of critical
importance for the subsequent computation of solution of the radiative
transfer problem for hydrogen.

This non-equilibrium EOS essentially solves the non-equilibrium rate
equations for hydrogen with prescribed radiative rates, together with
equations for energy, particle and charge conservation. By using
prescribed rates there is no need to solve the transfer equation, and
therefore the problem becomes computationally tractable in 3D
simulations.

We refer the reader to 
\citet{2007A&A...473..625L}
for more details about the non-equilibrium EOS, and to
\citet{2011A&A...531A.154G}. 
for an extensive description of \Bifrost.

Our simulation has a grid of $504 \times 504 \times 496$ grid cells,
with an extent of $24 \times 24 \times 16.8$\,Mm. In the vertical
direction the grid extends from 2.4\,Mm below to 14.4\,Mm above
average optical depth unity at 500\,nm, encompassing the upper
convection zone, photosphere, chromosphere and lower corona. The $x$
and $y$-axes are equidistant with a grid spacing of 48\,km. The
$z$-axis is non-equidistant. It has a grid spacing of 19\,km between
$z=-1$ and $z=5$\,Mm, while the spacing increases towards the upper
and lower boundaries to a maximum of 98\,km at the coronal
boundary. The magnetic field has a predominantly bipolar structure,
that manifests itself in the photosphere as two clusters of magnetic
concentrations of opposite polarity. The magnetic field was introduced
into a relaxed hydrodynamical simulation through specifying the vertical
magnetic field at the bottom boundary with a potential field extrapolation giving
the magnetic field throughout the box. The field at the bottom boundary was given 
with an averaged signed field of zero and two patches of opposite polarity 
separated by 8\,Mm. The simulation ran for 3000\,s before switching on
hydrogen non-equilibrium ionization. The snapshot we analyze is after  850\,s
with hydrogen non-equilibrium ionization.

The average unsigned magnetic field strength at $z=0$\,Mm in the
snapshot we selected is 50\,G. The snapshot is further characterized
in Fig.~\ref{fig:snapcharac}. The upper-left panel displays the
vertically emergent intensity in the continuum radiation bin. It shows
granules and some bright intergranular structures that correspond
to kG magnetic field concentrations as can be seen by comparing with
the upper-right panel, which shows the vertical magnetic field strength at
$z=0$\,Mm. The two clusters of opposite polarity field stand out in
black and white.
The lower left panel shows the gas temperature in the chromosphere at
a height of 1.7\,Mm. The corona dips down to below this height above
the photospheric magnetic field concentrations. Between the magnetic
patches the temperature shows filamentary structure that follows
the general orientation of the magnetic field. This can be seen in the
lower-right panel, that shows the angle of the magnetic field vector
with the horizontal, together with arrows pointing in the direction of
the horizontal component of the field, also at $z=1.7$\,Mm.

We performed the radiative transfer computations based on this snapshot
with the \multitd\ code
\citep{2009ASPC..415...87L}.
This is a radiative transfer code capable of handling 3D model
atmospheres. It can evaluate the radiation field for each column of
the atmosphere separately (column-by-column plane-parallel
approximation), or it can do so in full 3D, taking
the horizontal structure into account. 

\multitd\ employs the
accelerated lambda iteration method developed by
\citet{1992A&A...262..209R} 
with the extension to treat effects of partial frequency
redistribution using the angle-averaged approximation by
\citet{2001ApJ...557..389U}. 
In order to make the 3D radiative transfer problem computationally
tractable we halved the horizontal resolution of the MHD snapshot and
resampled the vertical grid to 200 equidistant points between $z=-0.5$\,Mm
and $z=5.5$\,Mm to cover just the formation range of the various
hydrogen transitions. For the radiative transfer computations we thus
had a horizontal resolution of 96\,km and a vertical resolution of
32\,km.

\edt{We used a 5-level plus continuum hydrogen model atom. For computations
assuming 1D plane-parallel geometry we used a 5-point Gauss-Legendre
quadrature to compute the angle integration for each ray direction (so
10 rays in total) for the radiation field. For full 3D computations we
employed the A4 set of 
\citet{carlson1963}
with 3 rays per octant, and thus 24 rays in total. \multitd\ does not
include an equation for charge conservation. The non-LTE proton
density we obtained is larger than the electron density at
certain locations in the model atmosphere.}

\section{Influence of PRD in Lyman lines on \Halpha} \label{sec:prdtransfer}

\begin{figure}
  \includegraphics[width=\columnwidth]{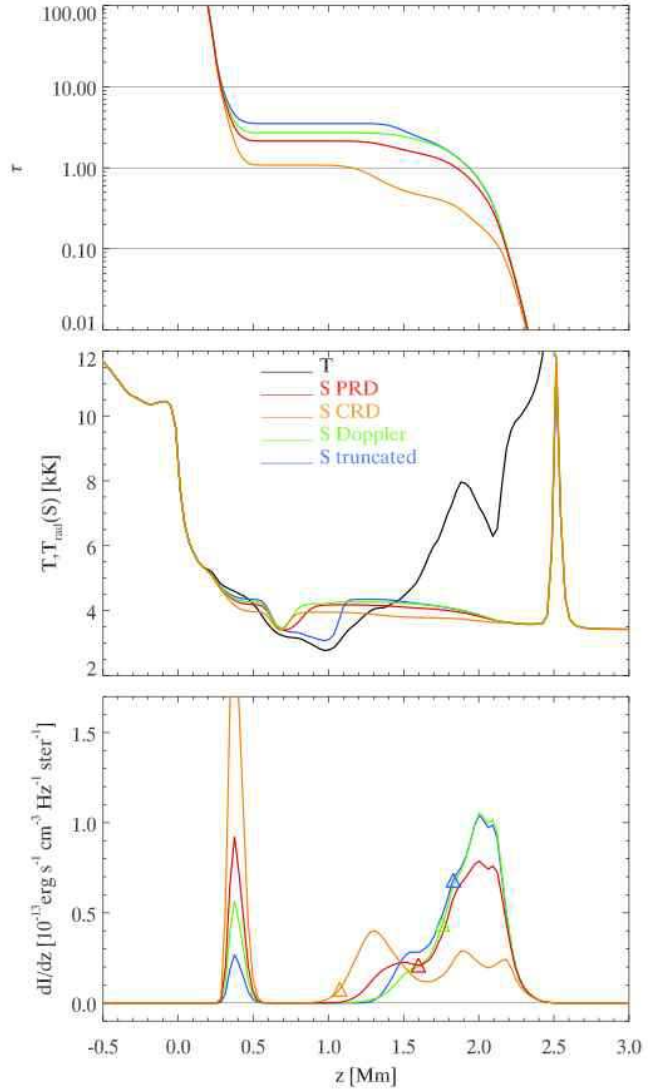}
  \caption{Comparison of \Halpha\ line formation between various
    methods to treat radiative transfer in the Lyman-$\alpha$ and
    $\beta$ lines in a column of a 3D model atmosphere. In all panels
    the red curve indicate results obtained with angle-averaged PRD,
    orange with CRD, green with a Doppler profile and blue with a
    truncated Voigt profile. Top: optical depth versus height, the
    thin horizontal lines indicate $\tau$ equals 0.1,1 and 10, to
    guide the eye.  Middle: gas temperature (black curve) versus
    height, with the total source function overplotted.  Bottom:
    contribution function to intensity of the \Halpha\ line center
    versus height. The thin horizontal line indicates $\rmd I / \rmd z
    = 0$ to guide the eye. The average formation height is indicated
    by the triangles. Emergent line profiles for this atmosphere are
    shown in Fig.~\ref{fig:prd-comp-prof}.}
  \label{fig:prd-comp}
\end{figure}

\begin{figure}
  \includegraphics[width=\columnwidth]{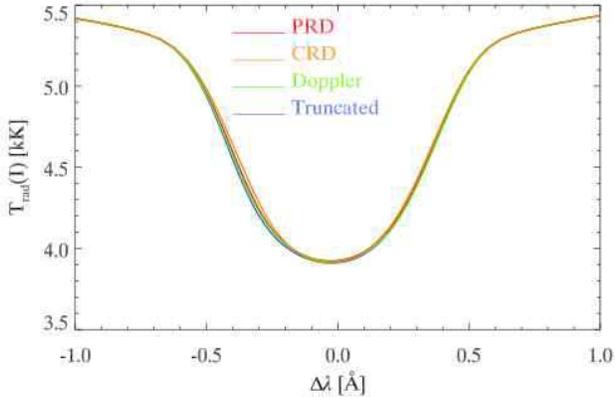}
  \caption{Comparison of the \Halpha\ line profile computed with the
    Lyman-$\alpha$ and $\beta$ lines treated with angle-averaged PRD (red),
    CRD (orange), a Doppler profile (green) and a truncated Voigt
    profile (blue). Details on how the line core is formed are shown in
    Fig.~\ref{fig:prd-comp}.
  \label{fig:prd-comp-prof}}
\end{figure}

\begin{figure}
  \includegraphics[width=\columnwidth]{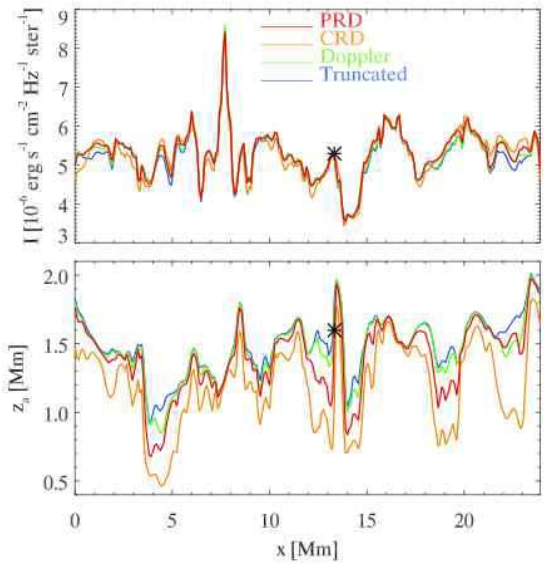}
  \caption{Comparison of the \Halpha\ line core formation with the
    Lyman-$\alpha$ and $\beta$ lines treated with angle-averaged PRD
    (red), CRD (orange), a Doppler profile (green) and a truncated
    Voigt profile (blue) in a 2D slice of a 3D radiation-MHD snapshot
    computed with \Bifrost. Top panel: line core intensity versus
    spatial location. Bottom panel: average formation height versus
    spatial location. The stars indicate the column used in
    Figs.~\ref{fig:prd-comp} and~\ref{fig:prd-comp-prof}
  \label{fig:prd-comp-ivsx}}
\end{figure}

The Lyman lines of hydrogen are strongly affected by partial
redistribution of photons over the line profile 
\citep[PRD, \eg][]{1973ApJ...185..709M,1981ApJS...45..635V}.
The Lyman line transfer affects the atomic level populations and
therefore also the formation of the \Halpha\ line. The computation of
PRD effects in a three-dimensional atmosphere is currently
unfeasible. This is partly due to the increase in computational cost
compared to using the simplifying assumption of complete redistribution of photons
over the line profile (CRD), and partly due to algorithm stability: the
strong gradients present in MHD models very often lead to
non-convergence of the $\Lambda$-iterations. We therefore investigated
the effect of various approximations to PRD effects in the Lyman
lines, in the hope of finding a simple recipe that has the same influence on the formation
\Halpha\ as PRD.

Inspired by 
\citet{1973ApJ...185..709M}
and
\citet{1977BAAS....9Q.432A}
we compared four different treatments of the Lyman-$\alpha$ and
$\beta$ lines: using angle-averaged PRD with a Voigt profile (hereafter
`PRD case'), CRD with a Voigt profile (`CRD case'), CRD with a Voigt profile
truncated at 48 km\,s$^{-1}$ away from line center (`truncated case')
and CRD with a Gaussian profile with Doppler broadening only (`Doppler case'). 

We took an $xz$-slice from the 3D \Bifrost\ snapshot at $y=0$\,Mm and
solved the SE radiative transfer problem for hydrogen with
\multitd\ for each of the four different approximations, but otherwise
identical input data. Each of the 252 columns was treated as a
plane-parallel atmosphere. It turned out that the differences between
the various cases showed behavior common to nearly all columns. We
therefore show the results for one representative column only in
Figs.~\ref{fig:prd-comp} and~\ref{fig:prd-comp-prof}.

The top panel of Fig.~\ref{fig:prd-comp} shows the optical depth as
function of height for the various cases. The CRD case shows the
lowest optical depth \edt{of the \Halpha\ core} at any given height, the truncated case the
highest; the Doppler case is closest to the PRD case. In all cases
there is no increase in optical depth \edt{with geometrical depth} around $z=0.8$\,Mm, indicating
the `opacity gap' typical of lines with an excited state as a lower
level
\citep{2006A&A...449.1209L}. 
This can be seen in observations when stepping through a series of
\Halpha\ images at successively larger wavelength offset from line
center, where the overall appearance jumps from high-chromospheric
structure to photospheric granulation, without intermediate reversed
granulation such as in the \CaII\ H line.

The ordering of the optical depth scales can be explained in terms
of the thermalization depths of the Lyman-$\alpha$ line, or in other
terms, as the probability in photon escape in the line wing
\citep[\cf][]{1978stat.book.....M}. 
The
Doppler and truncated case have a very weak or no line wing, and
therefore a low thermalization depth and few line photons escape
through the line wings. In contrast, the Voigt case has prominent
wings, a large thermalization depth and a fraction of the line photons
escape into space through the wings. Therefore the Lyman-$\alpha$ line
has a larger net downward rate and hence lower $n=2$ population and
hence lower \Halpha\ opacity in the Voigt case than in the Doppler and
truncated cases. The PRD case is intermediate: \edt{Lyman-$\alpha$ and Lyman-$\beta$} behave more like
Doppler lines in the upper chromosphere and more like a Voigt lines
deeper down in the atmosphere.

The middle panel \edt{of Fig.~\ref{fig:prd-comp}} shows a comparison of the gas temperature and the
total source function $S$ in the \Halpha\ line core.  The temperature
drops down from the photosphere to a minimum at $z=1.25$\,Mm, and then
slowly increases to transition-region temperatures, with a dip at
$z=2.0$\,Mm. All source functions show $S=B$ (with $B$ the Planck
function) up to $z=0.2$\,Mm after which they decouple from the local
temperature. The dips in the source functions at around $z=0.8$\,Mm are
caused by the very low line opacity causing the background source
function to dominate the total source function. The large spike at
$z=2.5$\,Mm is likewise caused by the background processes and plays
no role in the formation of the line. Except where the
background source function is appreciable, in the core-forming region
($0.3 < z < 2.5$\,Mm) the truncated and Doppler cases show a larger
source function than in PRD. The CRD case a lower source function. The
bottom panel shows the contribution function (CF) to intensity. All CFs
are multiple-peaked and very wide, being non-zero over a height range
of 2\,Mm. The CRD case has the lowest average formation height
($z_\rma$, indicated by a triangle), the truncated case the highest.

Fig.~\ref{fig:prd-comp-prof} shows the vertically emergent line
profile for the same column as shown in Fig.\ref{fig:prd-comp}. The
profiles are identical in the line wing. In the $\pm0.6$\,\AA\
around line center the differences are small but visible.

\begin{table}
\caption{Comparison of PRD approximations}
\label{table:prd-comp}
\centering
\begin{tabular}{c c c}
\hline\hline 
Case & $<I/I_{\mathrm{PRD}}-1>$ & $<z_\rma-z_{\mathrm{a,PRD}}>$ (km)
\\
\hline 
Voigt & 0.026 & -203 \\
Truncated & 0.019 & 96 \\
Doppler & 0.011 & 65 \\
\hline
\end{tabular} 
\end{table}

The top panel of Fig.~\ref{fig:prd-comp-ivsx} compares the variation
of the line-core intensity and average formation height for all
columns in the $xz$-slice. The variations of the \edt{\Halpha}\ intensity due to the
various PRD approximations in \edt{Ly-$\alpha$\ and Ly-$\beta$\ are} much smaller than the variations due to
the different atmospheric properties. Table~\ref{table:prd-comp}
compares the relative difference in line core intensity between the
various approximations and the PRD case. The Doppler case gives the
best approximation to the PRD case, with on average only 1.1\%
difference in line-core intensity.

The bottom panel similarly compares the average formation height. Here
the variations caused by the various PRD approximations are larger. The
truncated case yields the highest $z_\rma$, followed by the Doppler
case, the PRD case, and lowest, the CRD case. As can be seen in
Table~\ref{table:prd-comp}, the Doppler case approximates the PRD case
best, with on average a 64\,km larger formation height. The
differences between the Doppler and PRD cases are large where
$z_\rma$ is small, for example at $x=4$\,Mm in the lower panel
of~\ref{fig:prd-comp-ivsx}; they are small where $z_\rma$ is large, such
as at  $x=1$\,Mm.

From the above analysis we conclude that it is possible to use Doppler
profiles in the Lyman-$\alpha$ and $\beta$ lines without significantly
influencing the formation of \Halpha.

\section{The influence of 3D transfer on \Halpha} 
\label{sec:3Dtransfer}

\begin{figure*}
  \includegraphics[width=\textwidth]{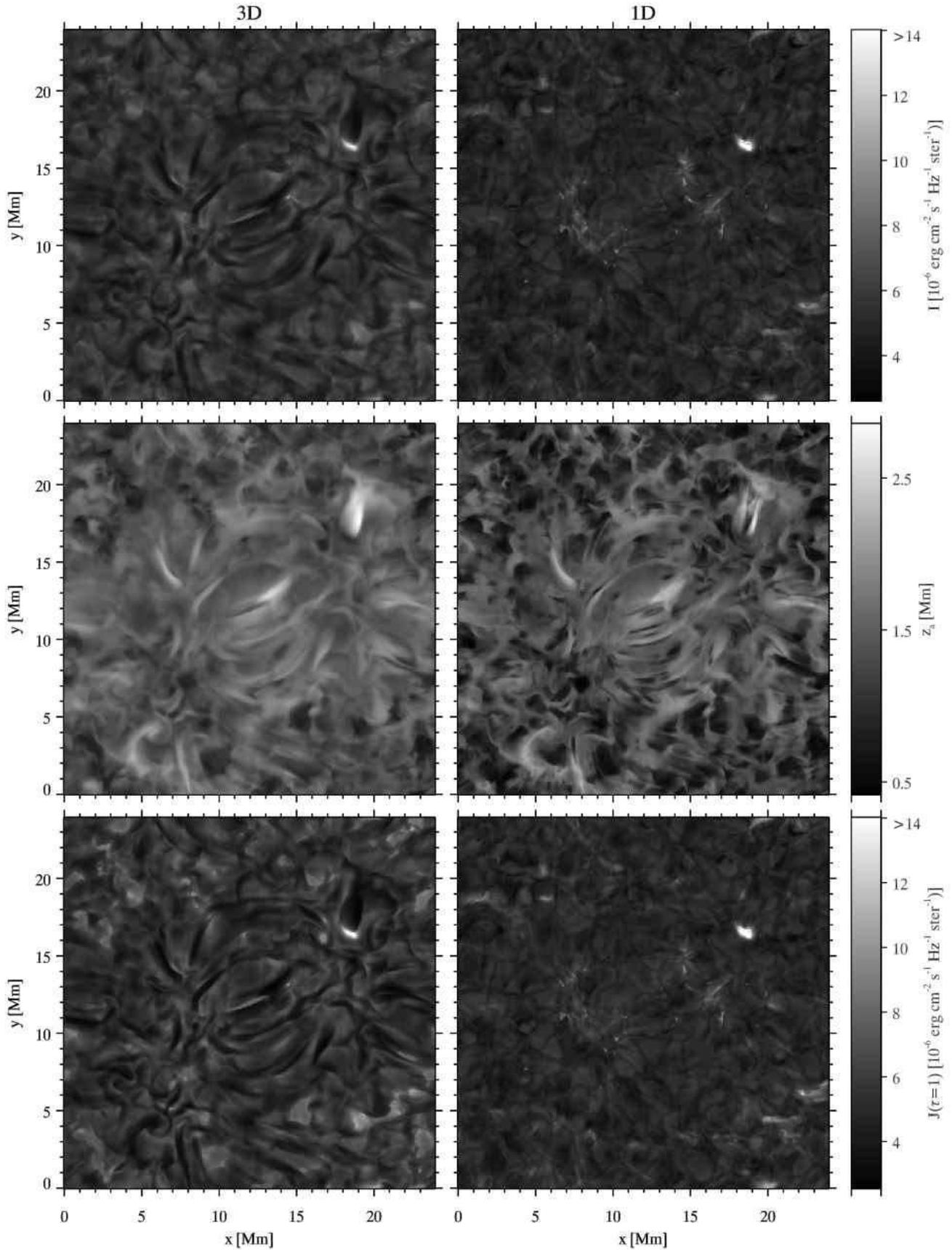}
  \caption{Comparison of \Halpha\ line-core formation between full 3D
    radiative transfer (left-hand column) and treatment of each column
    in the snapshot as a plane-parallel atmosphere (right-hand
    column). First row: vertically emergent intensity. Second row:
    average formation height. Third row: angle-averaged radiation
    field at optical depth unity. }
  \label{fig:ha-1d-3d}
\end{figure*}

The conclusions of Sec~\ref{sec:tdtransfer} and~\ref{sec:prdtransfer},
namely that it is possible to model \Halpha\ line formation assuming
statistical equilibrium and approximating PRD effects in the Lyman
lines, makes it numerically feasible to perform 3D radiative
transfer. Effects from 3D radiative transfer are expected to be
important for strongly scattering lines like \ion{Na}{1}\,D$_{1}$ 
\citep{2010ApJ...709.1362L} 
but not so important for lines that have a strong coupling of the
source function to the temperature like the \CaII\ infrared triplet
lines
\citep{2009ASPC..415...87L}. 
The \Halpha\ line is strongly scattering as can be seen in the
interesting comparison of \CaII\ 8542 to \Halpha\ formation in Fig.~8 of 
\citet{2009A&A...503..577C}. 
The \Halpha\ source function decouples from the temperature already in
the photosphere in the 1D FAL model C of
\citet{1993ApJ...406..319F} 
and is afterwards essentially equal to the angle-averaged radiation field.
An important corollary of this is that the line-core intensity is
determined mainly by the photospheric temperature, \ie the line-core
response function to temperature peaks in the photosphere
\citep{2004ApJ...603L.129S,2006ASPC..354..313U}. 

We therefore solved the radiative transfer problem for hydrogen in our
entire \Bifrost\ snapshot twice, once with full 3D transfer, and once
treating each column as a plane-parallel atmosphere (hereafter called
1D transfer). We assumed SE and the approximation to PRD from
Sec.~\ref{sec:prdtransfer}. No microturbulence was added. The results
are shown in Fig.~\ref{fig:ha-1d-3d}.

The top row compares the vertically emergent intensity between 3D and
1D. They show a starkly different scene. The 3D computation shows
striking, predominantly dark, fibril-like features emanating from the
photospheric field concentrations, which appear to follow the magnetic
field lines (compare to Fig.~\ref{fig:snapcharac}).  They are
superimposed on a background of irregularly shaped patches of roughly
2\,Mm diameter of varying brightness. The 1D computation instead shows
a distinctly photospheric scene with granulation and magnetic bright
points, much more reminiscent of observations in the line wing.

The second row shows the average formation height. The 3D image shows
that the structures that appear dark in the line-core image are formed
highest. The irregular background in the intensity image has a lower
$z_\rma$. In the 1D case the situation is largely the same, except
that it shows more small scale structure (due to the absence of
smoothing 3D radiation effects on the opacity) and the patches of
small $z_\rma$ away from the magnetic concentrations.

The bottom row shows the angle-averaged line-core radiation field
$J_{\nu0}$ at the local optical depth unity. Both the 1D and 3D images
appear very similar to the corresponding intensity images. This
validates the Eddington-Barbier (EB) approximation
$I_{\nu0}(\mu) \approxeq S_{\nu0}(\tau=\mu)$ for vertical rays ($\mu=1$), and also
that the core source function is set by scattering ($S_{\nu0}(\tau=1) \approxeq J_{\nu0}(\tau=1)$).

 This also explains why the 1D intensity image shows granulation and
 bright points. The \Halpha\ source function at optical depth unity is
 given by the $J_{\nu0}(\tau=1)$, which in turn is \edt{mainly} set in the
 photosphere. In 1D there is no smoothing of the radiation field, and
 $J_{\nu0}(\tau=1)$ in a given column correlates strongly with
 temperature at the thermalization depth in that column. This
 temperature is lowest in intergranular lanes, higher inside granules
 and highest in magnetic elements. The resulting image therefore shows
 the structure of the photosphere, even though optical depth unity
 lies much higher.

We conclude that it is essential to include 3D radiative transfer
effects in the formation of the \Halpha\ line core. In the next section
we analyze the results of our 3D computation in more detail.

\section{Analysis of the 3D computation}     \label{sec:analysis}

\subsection{\Halpha\ core intensity correlates with average formation
  height} \label{subsec:icorza}

\begin{figure}
  \includegraphics[width=\columnwidth]{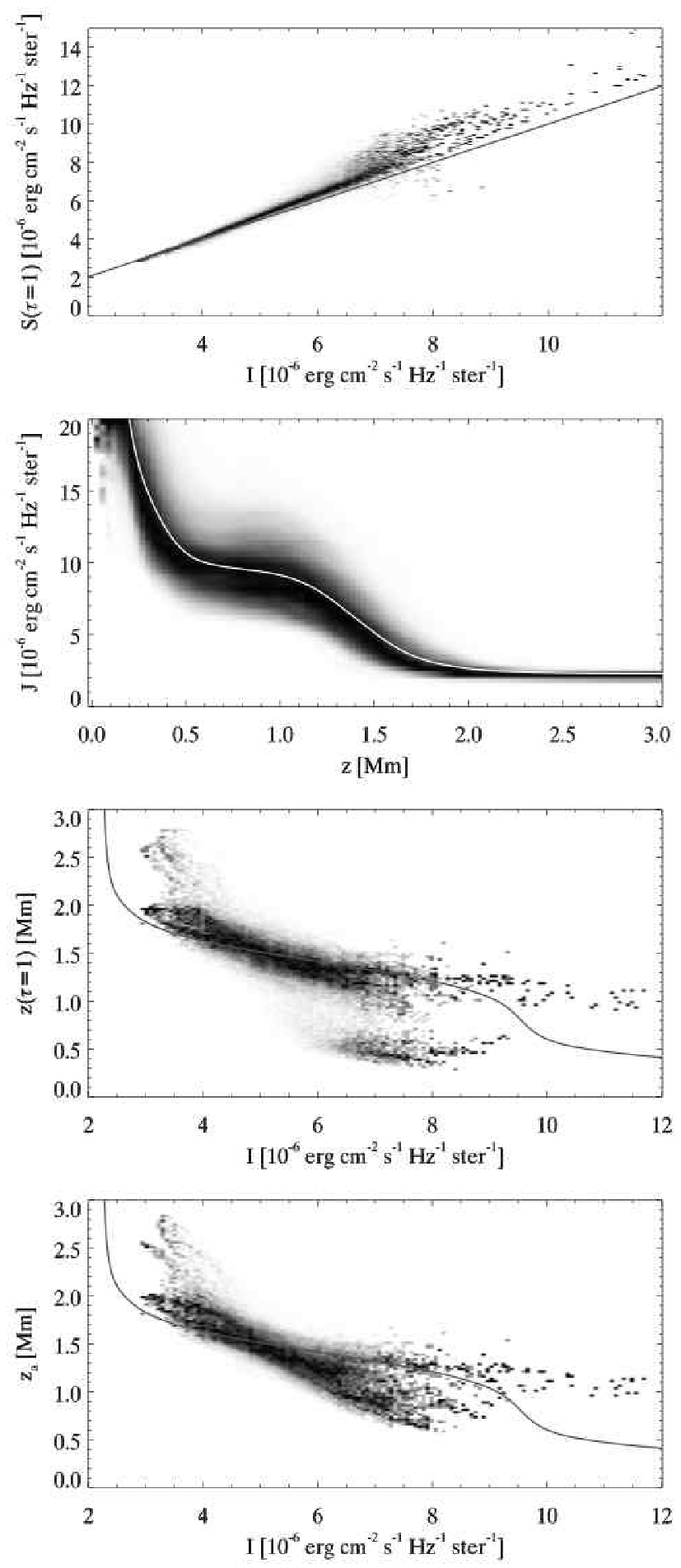}
  \caption{Panels, from top to bottom: (1) Probability density
    function (PDF) of the source function at optical depth unity as
    function of emergent \Halpha\ intensity in the line-core. The grey line indicates
    $S=I$.  (2) PDF of the angle-averaged radiation field $J$ as
    function of height. The white curve shows the horizontal average
    of $J$ as function of height.  (3) PDF of the line-core optical
    depth unity height as function of emergent \Halpha\ intensity. The
    gray curve displays the same quantity as the white curve from
    panel 2, but transposed to account for the different arrangement of
    the axes. (4)
    PDF of the average formation height as function of emergent
    \Halpha\ intensity. The gray curve is the same as for panel 3. All panels have each column individually scaled
    to maximum contrast for improved readability.
  \label{fig:corr-za-ha}}
\end{figure}

Fig.~\ref{fig:ha-1d-3d} suggests a correlation between
\Halpha\ intensity and average formation height for the 3D computation. In
Fig.~\ref{fig:corr-za-ha} we investigate the cause of this effect.
The top panel shows the probability density function (PDF) of the
line-core source function at optical depth unity as function of
emergent \Halpha\ intensity in the line core. It confirms that the Eddington-Barbier
relation is valid.  The second panel shows the PDF of $J_{\nu0}$ as
function of height. There is a rather small spread of $J_{\nu0}$
around the horizontal average of the angle-averaged radiation field
$<J_{\nu0}>_{xy}$. The latter is a decreasing function of height up to
2\,Mm height, after which it becomes flat.

The third panel shows the PDF of the line core optical depth unity
height as function of emergent intensity. There is a strong
correlation between the two quantities between $I$=4$\times$10$^{-6}$
and $I$=7$\times
$10$^{-6}$\,erg\,cm$^{-2}$\,s$^{-1}$\,Hz$^{-1}$\,ster$^{-1}$. At lower
intensity the correlation shows two protrusions with a steeper slope.
Between $I$=7$\times$10$^{-6}$ and $I$=8$\times$
10$^{-6}$\,erg\,cm$^{-2}$\,s$^{-1}$\,Hz$^{-1}$\,ster$^{-1}$ the
distribution is bimodal, with maxima at around $z=0.5$\,Mm and
$z=1.3$\,Mm. This bimodality is caused by the mid-chromospheric
opacity gap: this gap is typically located around optical depth unity
for those emergent intensities. Depending on whether $\tau=1$ is
reached above or below this gap, a given column in the atmosphere
contributes to the low or the high cluster of points in the PDF. 

At even higher emergent intensities the $\tau=1$ height appears
independent of the intensity. The number of columns with this
intensity is small, so it is unclear whether this is significant.
The bimodality can be largely removed by using the average formation
height instead of the height of optical depth unity. The former is
much less sensitive to the location of the $\tau=1$ height than the
latter. This is shown in the bottom panel.

The correlation between I and $z_\rma$ (or $z(\tau=1)$) can now easily
be explained: The Eddington-Barbier approximation is valid, and
scattering ensures $S_{\nu0}=J_{\nu0}$. The radiation field $J_{\nu0}$
shows relatively little variation with horizontal position and its
horizontal average is monotonically decreasing with height. Therefore,
larger average formation height leads to lower intensity. Above
$z=1.7$\,Mm the radiation field is approximately constant with height
and the formation height and emergent intensity are no longer
correlated.

\subsection{\Halpha\ line-core width correlates with temperature} \label{sec:lw_tg}

\begin{figure}
  \includegraphics[width=\columnwidth]{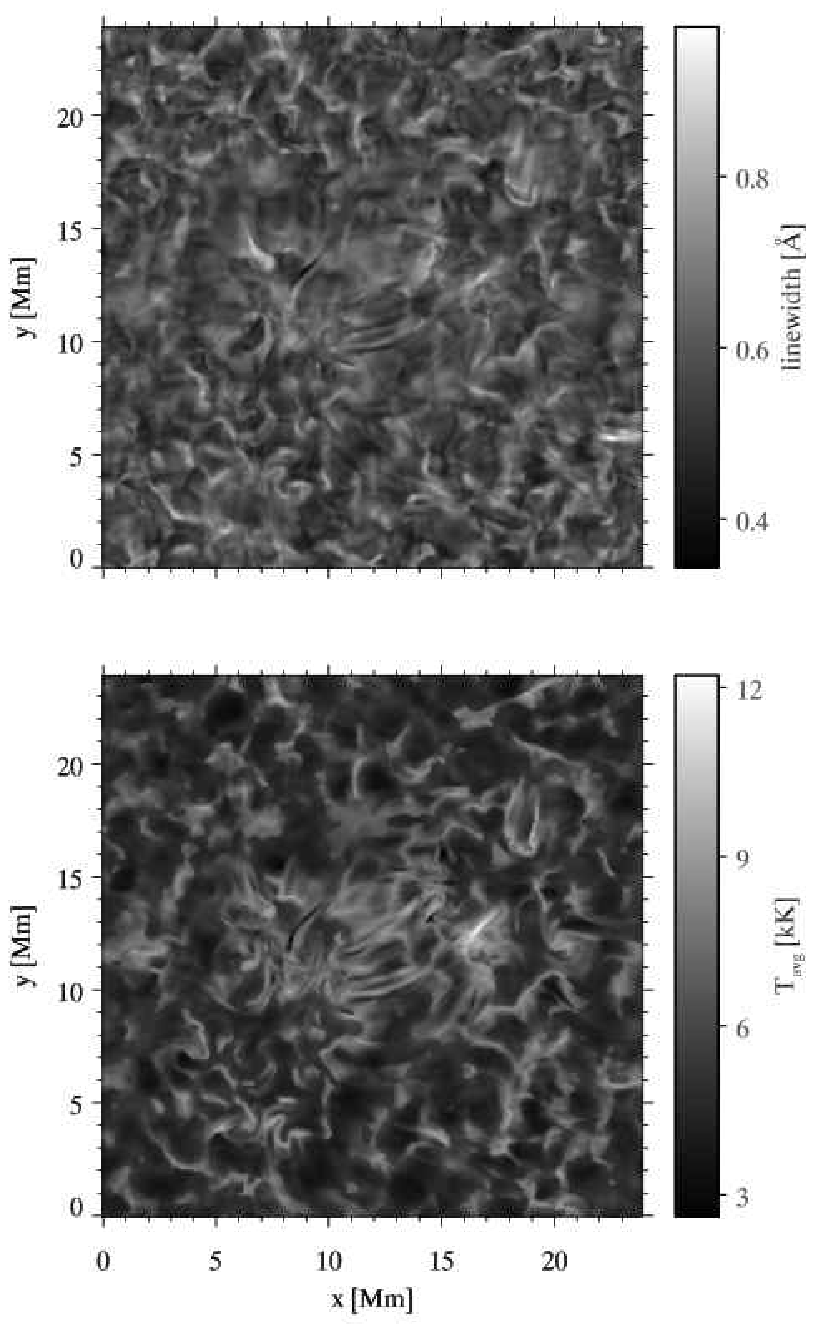}
  \caption{Comparison of line-core width (top) and temperature averaged
    between $\tau=0.5$ and $\tau=5$ at the wavelength of the profile
    minimum (bottom).
  \label{fig:ims-lw-tg}}
\end{figure}

\begin{figure}
  \includegraphics[width=\columnwidth]{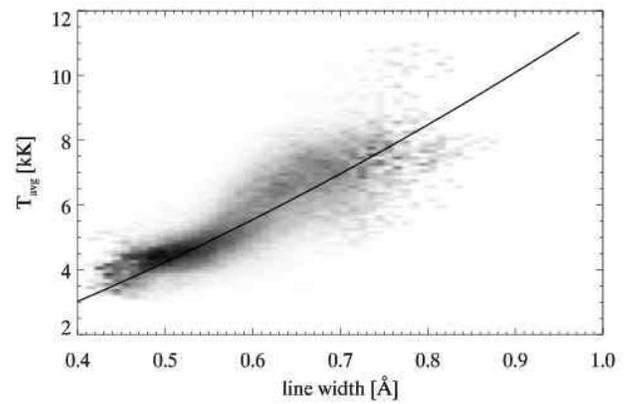}
  \caption{Probability density function (PDF) of the temperature
    averaged between $\tau=0.5$ and $\tau=5$ at the wavelength of the
    profile minimum against the line-core width, after smoothing with
    a $3 \times 3$ moving boxcar average. The black curve is a
    quadratic fit to the data.
  \label{fig:cor-lw-tg}}
\end{figure}

Because of the low mass of the hydrogen atom it is expected that the
width of the \Halpha\ line core is sensitive to the temperature in
the line forming region. Because the Doppler speed ($\sqrt{2kT/m_{\rmH}}=13$\,
km\,s$^{-1}$ for $T=10$\,kK) is comparable to or larger than typical gas velocities
in the chromosphere, the core width is not very sensitive to velocity
gradients along the line-of-sight in the line forming region. 

Because we are interested in a diagnostic of the chromospheric
temperature we choose to measure the line width close to the line
core. For each pixel in our synthetic image we determined the profile
minimum and maximum ($I_\mathrm{min}$ and $I_\mathrm{max}$) and then
defined the core width as the wavelength difference between the
intersections of the profile and the line $I=I_\mathrm{min}+0.1
(I_\mathrm{max}-I_\mathrm{min})$. We then computed for each pixel the
temperature averaged over the height range between $\tau=0.5$ and
$\tau=5$ at the wavelength of the profile minimum. The height range
was extended to larger optical depths because one looks deeper in the
atmosphere at the wavelengths that define the line-core width than in
the line-core.

The results are shown in Fig.~\ref{fig:ims-lw-tg}. Comparison of the
two panels shows that there is a decent correspondence between large
width and large average temperature, but in general no clear
correspondence for pixels with low width and low temperature. The darker
areas in  the upper panel show a granulation pattern (compare with
Fig.~\ref{fig:snapcharac}), while the average temperature panel does not.
The notable exception are the dark elongated structures which clearly
show up near the center of both panels. Fig.~\ref{fig:cor-lw-tg} shows
the PDF of the average temperature versus the core width after
smoothing both with a $3 \times 3$ moving boxcar average. The
averaging serves to remove outliers in the correlation.

This figure confirms the impression obtained from the images: the two
quantities are quite well correlated at higher values, but in the low
to intermediate regime (below a line width of 0.6\,\AA) the correlation
is rather weak.

The correlation can be understood using optically thick line
formation,the Eddington-Barbier approximation and the fact that the
run of source function with height is monotonically decreasing and
rather similar for each column
(\cf\ Fig.~\ref{fig:corr-za-ha}). Because Eddington-Barbier is valid,
one looks deeper and deeper into the atmosphere as one moves away from
the line center and one sees a higher and higher source function and
hence emergent intensity. A low temperature means a narrow absorption
profile, hence steep gradient of the height of optical depth unity
with wavelength and therefore a narrow line. At higher temperature the
absorption profile is wider, a less steep $\rmd z(\tau=1) / d\lambda$
gradient and therefore a wider line. The actual line width at any
given location therefore depends on both the temperature, and the
gradient of the source function with optical depth in the line forming
region. Variation in the latter is the main cause of the scatter in
Fig.~\ref{fig:ims-lw-tg}.

\subsection{No correlation of \Halpha\ opacity with temperature in the
  upper chromosphere} \label{sec:cor_tg_n2nt}

\begin{figure}
  \includegraphics[width=\columnwidth]{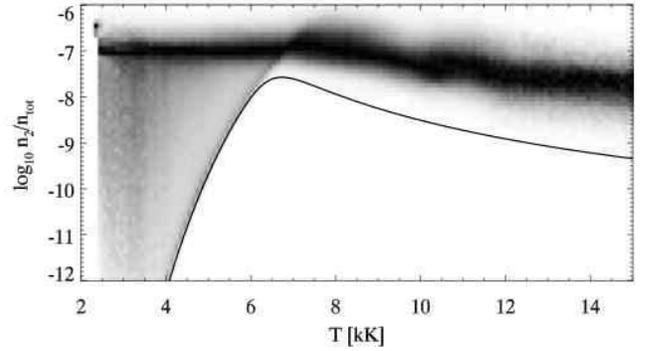}
  \caption{PDF of the relative occupation of the \Halpha\ lower level
    as function of the gas temperature at heights between 1\,Mm and
    3\,Mm. The black curve indicates the Saha-Boltzmann occupation for
    the average electron density over this height range,
    3.65$\times$10$^{10}$\,g\,cm$^{-3}$. Each column has been scaled
    to maximum contrast.
  \label{fig:cor-tg-n2_ntot}}
\end{figure}

\begin{figure}
  \includegraphics[width=\columnwidth]{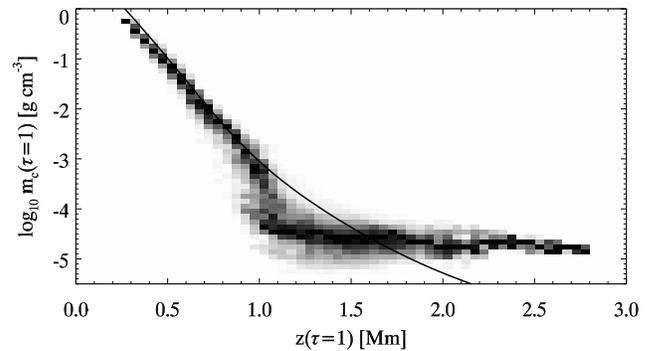}
  \caption{PDF of the column mass at the \Halpha\ line-core $\tau$=1 height
    as function of the $\tau$=1 height. Each column has been scaled
    to maximum contrast. The black curve represents the horizontal
    average of the column mass.
  \label{fig:cor-zt1-cmass}}
\end{figure}

The \Halpha\ line opacity is often considered to be sensitive to the
temperature. This is true in the photosphere and lower chromosphere where
the level populations are close to LTE, but does not hold in the upper
chromosphere. There the \Halpha\ opacity per gram is determined by
the ionization degree
\citep{2002ApJ...572..626C}
and the radiation field. The ionization degree is insensitive to
temperature variations over time due to the long
ionization-recombination timescale. The radiation field is non-locally
determined and therefore independent of the local temperature.

We demonstrate this in Fig.~\ref{fig:cor-tg-n2_ntot}. It shows the $n$=2 population
relative to the total (neutral plus ionized) hydrogen density for each
grid point in our simulation with a height between 1\,Mm and 3\,Mm. At
temperatures below 4\,kK the distribution is very wide, at higher
temperatures the distribution narrows considerably and its peak value
varies little over the whole temperature range. The black curve
indicates the LTE value of the relative occupation number, which
varies by many orders of magnitude.

The temperature-insensitivity of the upper-chromospheric opacity means
the opacity is mainly determined by the mass density. Therefore, the
optical depth scale in the upper chromosphere is proportional
to the column mass. This is shown in Fig.~\ref{fig:cor-zt1-cmass},
where we show the PDF of the column mass at the $\tau$=1 height as function of
the $\tau$=1 height. It shows that if the $\tau$=1 height is above
1.2\,Mm, it has a constant column mass of roughly
3$\times$10$^{-5}$\,g\,cm$^2$. 

At lower heights the opacity is sensitive to both the temperature and
the mass density. The column mass at the lower lying $\tau$=1 heights
is therefore no longer constant, and is close to the horizontal
average of the column mass instead.

\subsection{Fibril-like structures align with the magnetic field in a low  $\beta$ environment}

\begin{figure}
  \includegraphics[width=\columnwidth]{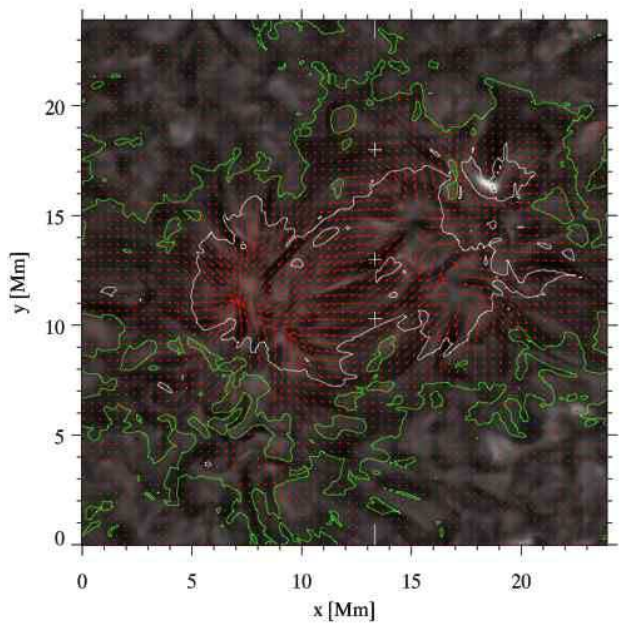}
  \caption{Image of the line-center intensity. Red arrows indicate the
    direction of the horizontal component of the magnetic field at the
    average formation height. Arrow length is proportional to field
    strength. The contours indicate values of plasma $\beta$ equals
    0.01 (white) and 0.1 (green) at the average formation height. The
    white tick marks indicate the cut used in
    Fig.~\ref{fig:yzslice}. White crosses indicate locations of
    fibrils crossing this cut. This figure is best viewed by magnifying
    the PDF file on a computer screen.
  \label{fig:btrace}}
\end{figure}

Chromospheric fibrils in \Halpha\ are associated with the presence of
photospheric magnetic elements and are conventionally thought to trace out
magnetic field lines, but this has so far only been shown for the
\CaII\ 8542\,\AA\ line
\citep{2011A&A...527L...8D} 

Fig.~\ref{fig:btrace} investigates the alignment of the magnetic field
in the model atmosphere with the dark fibril-like structures in the
\Halpha\ line-core. The figure shows the emergent intensity with the
horizontal magnetic field direction and the plasma $\beta$ (ratio of
gas pressure over magnetic pressure) at the
average formation height overplotted. The fibrils are mainly located
within the $\beta=0.01$ contour, between the photospheric field
concentrations of opposite polarity. These fibrils are closely
following the magnetic field direction. Outside the $\beta=0.1$
contour there are virtually no fibril-like structures and the image
shows a more grainy appearance. Note that the semi-transparent fibrils
at $(x,y)=(13.3,18)$ Mm (around the upper white cross, best seen in
the upper-left panel of Fig.~\ref{fig:ha-1d-3d}) also align with the
field.

\subsection{Fibril-like structures caused by field-aligned density
  ridges}

\begin{figure*}
  \includegraphics[width=\textwidth]{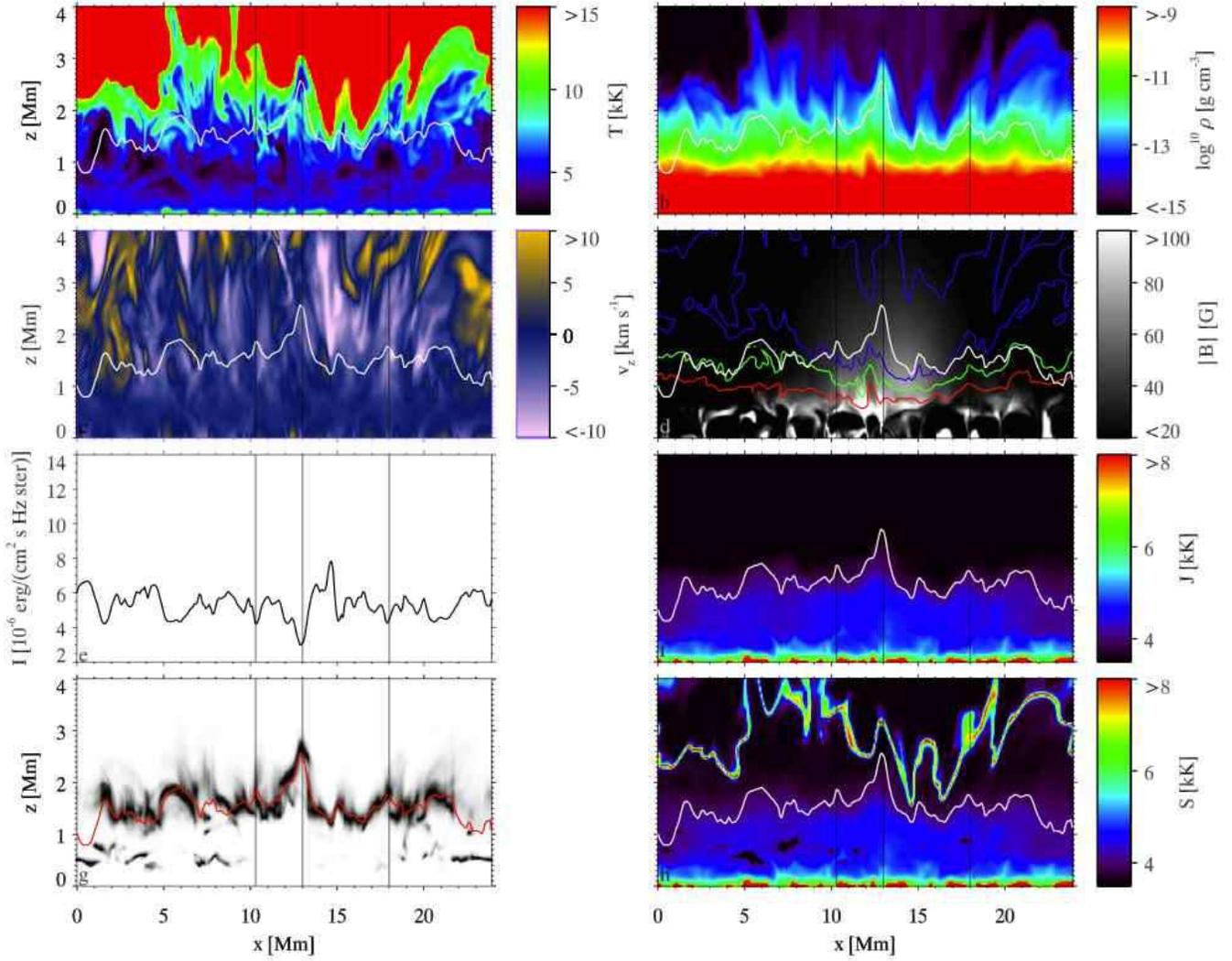}
  \caption{Vertical slice through the atmosphere along the cut
    indicated in Fig.~\ref{fig:btrace}. Vertical black lines indicate
    the positions of dark fibrils marked with white crosses in
    Fig.~\ref{fig:btrace}. The white or red curve indicates the
    average formation height. a: gas temperature; b: mass density; c:
    vertical velocity; d: magnetic field strength. The red, green and
    blue contours indicate plasma $\beta$ equals 1,0.1 and 0.01
    respectively. ; e: emergent line-center intensity f:
    angle-averaged line-center radiation field; g: contribution
    function to line-center intensity, each column scaled to maximum
    contrast; h: line-center source function.
  \label{fig:yzslice}}
\end{figure*}

\begin{figure*}
  \includegraphics[width=\textwidth]{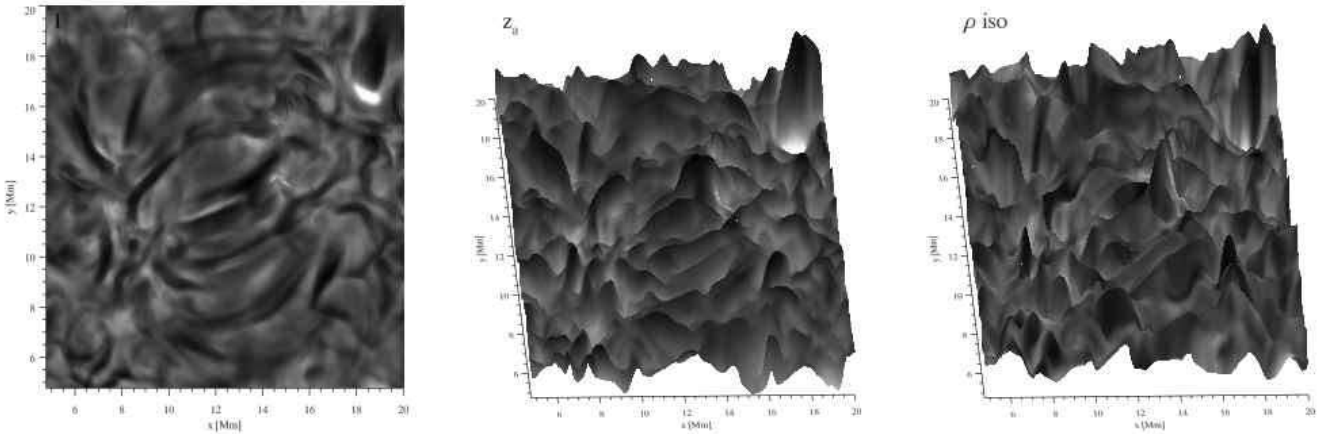}
  \caption{Display of ridge-like dark fibrils. Left panel: emergent
    intensity in the \edt{central part} of the atmosphere. Middle panel: 3D
    surface plot of the average formation height, with the color
    coding indicating the vertically emergent intensity. Right panel:
    iso-surface of the mass density at $\rho = 10^{-11}$\,g\,cm$^{-3}$,
    with the color coding again indicating the vertically emergent
    intensity. The dark fibrils follow ridges of enhanced
    chromospheric mass density.
  \label{fig:isorho}}
\end{figure*}

We now further investigate the cause of the dark structures by showing
an $yz$ slice through the atmosphere along the cut indicated in
Fig.~\ref{fig:btrace}. The panels a--d of Fig.~\ref{fig:yzslice} show the temperature, mass
density, vertical velocity and magnetic field strength. The fibril
locations (indicated by the vertical lines) do not show a particular
correlation with temperature and vertical velocity. Panel d again
shows they are located in an area with high magnetic field and low
plasma $\beta$. The mass density panel clearly shows that the fibrils
are located where there is enhanced chromospheric density, which
causes the average formation height to be larger and thus the emergent
intensity to be lower (see Sec.~\ref{subsec:icorza}). 

Panels e--h show the radiative transfer properties of the slice. The
fibrils appear as local minima in the intensity. In the core-forming
region the source function is nearly equal to the angle-averaged
radiation field. 

The contribution functions show a great variation in height of
formation, exhibit multiple peaks and typical extended upward streaks.
Away from the regions with strong
field the average formation height tends to be lower, with a strong
contribution from around 0.5\,Mm height. In regions with stronger
fields, there is no pronounced contribution from
this depth in the atmosphere.

Fig.~\ref{fig:isorho} shows the relation between \Halpha\ intensity,
average formation height and chromospheric mass density structure. The
left panel shows the emergent line-core intensity in a subfield of the
snapshot that shows fibrils. The middle panel shows a 3D rendering of
the average formation height surface, with the emergent intensity as
the image on the surface. It shows the dark fibrils as ridges with a
large average formation height. Finally, the right panel shows a 3D
representation of the iso-surface of the mass density at $\rho =
10^{-11}$\,g\,cm$^{-3}$, again with the emergent intensity as the
image on the surface. Wherever the average formation height shows a ridge,
there is a corresponding height increase in the density iso-surface.
Even though the exact shapes are not identical,
it is obvious that the dark fibrils between the two clusters of
opposite polarity field are caused by ridges of enhanced chromospheric
mass density. The fibrils that point away from both magnetic field
concentrations are also co-spatial with locations of increased mass
density, but because they are shorter, they appear more as mountains
than as ridges.

\section{Comparison with observations}

\begin{figure*}
  \includegraphics[width=\textwidth]{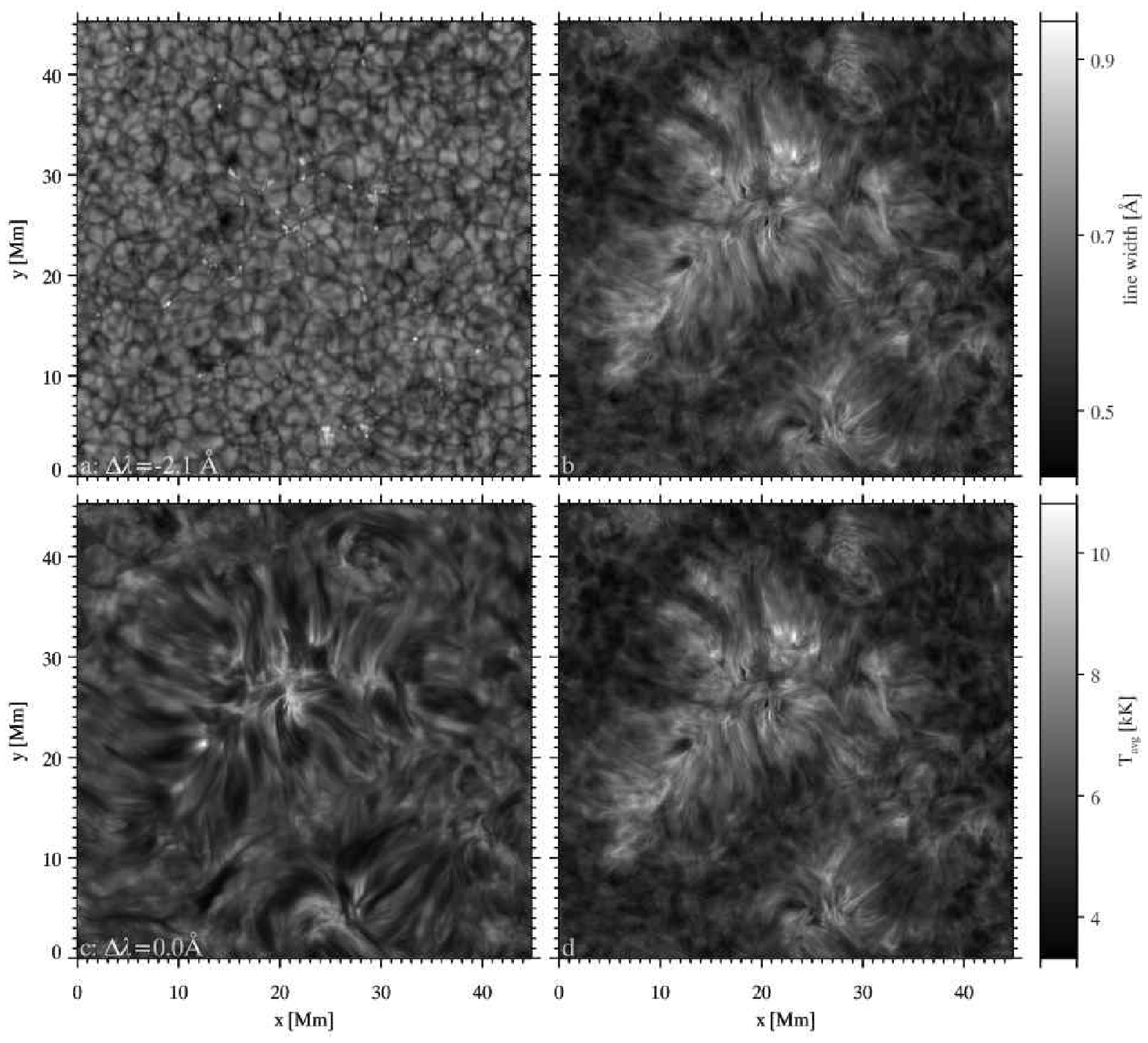}
  \caption{\Halpha\ observations obtained with the Swedish Solar
    Telescope. a: intensity in the blue wing at 2.1\,\AA\ from line
    center; b: line-core width; c: line-center intensity; d: average gas
    temperature in the formation height range as determined from the
    fit of Fig.~\ref{fig:cor-lw-tg}.
  \label{fig:obs}}
\end{figure*}

Fig.~\ref{fig:obs} shows observations of a quiet sun region taken in
the \Halpha\ line. The observations were obtained with the CRisp Imaging
SpectroPolarimeter
\citep[CRISP,][]{2008ApJ...689L..69S} 
at the Swedish 1-m Solar Telescope
\citep[SST,][]{2003SPIE.4853..341S} 
on La Palma on May 5,2011 at 7:53\,UT. 
%
The \Halpha\ spectral line was sampled with 35 line positions, from
$-$2.064~\AA\ to $+$1.290~\AA, and equidistant 86~m\AA\ stepping between
$\pm$1.290~\AA.
The time to complete a full line scan was 8~s.
The target area was a quiet 60$\times$60\arcsec region at disk center
centered at $(x,y)=(19$\arcsec,$5$\arcsec$)$.
The seeing conditions were very good and the data quality benefited
from the SST adaptive optics system
\citep{2003SPIE.4853..370S} 
and Multi-Object Multi-Frame Blind Deconvolution
\citep[MOMFBD,][]{2005SoPh..228..191V}
image restoration.
For more details on the data processing, we refer to
\citet{2009ApJ...705..272R}. 

Panel a shows the blue-wing image. It shows granulation and bright
points associated with magnetic concentrations. Panel c shows the
line-center image. There are two clusters of fibrils centered at
(27,7)\,Mm and (23,27)\,Mm pointing away from the photospheric bright
points. There appear to be no fibrils connecting the clusters, so it
is likely the magnetic field in the image is largely unipolar. At
larger distances from the bright points the line-core shows a more
grainy appearance without the elongated structures. This is seen most
clearly in the upper-left corner and close to the edge of the image at
the right-hand side.

The synthetic line-core image in the upper left panel of
Fig.~\ref{fig:ha-1d-3d} shows a similar structure, it shows
fibril-like structure close to the photospheric magnetic
concentrations and grainy structure elsewhere. The simulations lack
the dense packing of the fibrils in the observations and shows
typically shorter fibril-like structures. The width of the dark
fibrils in the observations are similar to the width of the dark
fibril-like structures in the simulations.

Panel b shows the line-core width, determined in the same way as for
the simulations (see Sec.~\ref{sec:lw_tg}). 
The fibrils show largest
width. The areas with lower width do not show a granulation pattern,
instead it shows a pattern reminiscent of reversed granulation or
acoustic shocks. Comparison of panel b and Fig.~\ref{fig:ims-lw-tg})
shows that the synthetic fibril-like structures show up much less in the
line-width measurement than the observed fibrils. The signature of the
granulation pattern in the simulations shows that the simulated
internetwork has a lower \Halpha\ opacity than the sun. The range of line-core
widths in the simulation is very similar to the one measured in the
observations.

Panel d finally shows the gas temperature derived from the core width
assuming the fit from Fig.~\ref{fig:cor-lw-tg}. It shows temperatures
between 4 and 8\,kK for internetwork areas.
The fibrils have a higher temperature of around 10\,kK. Note that
the brightest patches in the line-core image do not correspond to the
largest core width or derived temperature.

\section{Discussion and Conclusions} \label{sec:discussion}

We have presented results of a radiative transfer computation of the
\Halpha\ line based on a model atmosphere computed with a
state-of-the-art radiation-MHD code. This is the first time that a
simulation reproduces chromospheric fibril-like structures in synthetic
observables. 

The crucial ingredient for the formation of fibril-like structures is the full 3D evaluation of the transfer equation. The \Halpha\ source function decouples from the temperature already in the photosphere and the source function at optical depth unity in the chromosphere is therefore set by radiation emitted in a large volume and hence fairly smooth.

In our simulation the emergent line-core intensity correlates with the
average formation height: the lower the intensity, the higher in the
atmosphere the photons are originating. 
It is a very different
explanation than the one that is typically assumed by cloud modeling,
where variation in emergent intensity is assumed to be caused by
varying thermodynamic properties in a cloud that modulate the intensity emitted by the photosphere below.

We expect that this intensity-formation-height correlation is also
valid on the sun, as it is based only on the assumption that the \Halpha\
line is strongly scattering within its formation height range, so
that the source function is set by scattering and the angle-averaged
radiation field shows relatively little variation with horizontal spatial
position. However, the shape of the correlation curve depends on the
exact run of the source function with height. The latter depends on
the temperature at the thermalization depth and the chromospheric
opacity. These will be different in, for example, internetwork, network
or plage and different again in our simulation. When comparing intensities for small areas on the sun one can conclude that the lower the intensity the higher in the atmosphere one looks, but intensities in very different regions cannot be directly compared.

A largely similar argument holds for the correlation between line-core
width and gas temperature. This correlation can be simply explained by
optically thick line formation and the Eddington-Barbier approximation
and is therefore doubtlessly valid on the sun too. The exact shape of
the correlation depends on the variation of source function and
opacity with height. The fit shown in Fig.~\ref{fig:cor-lw-tg} is
therefore likely not universally valid. 

In addition, the method is sensitive to velocity gradients that are of
the same order as the line width. Such a gradient causes an increase
in line width compared to an identical atmosphere without a velocity
field. For example, the asymmetric profiles of `rapid blueshifted
events'
\citep{2009ApJ...705..272R} 
would show up as having increased temperature using the core-width measure, without this necessarily being
the case.

Still, if we assume it is valid and apply it, our observations yield
internetwork temperatures compatible with radiation-MHD simulations of
the internetwork chromosphere
\citep{1995ApJ...440L..29C,2000ApJ...536..465S,2004A&A...414.1121W,2007A&A...473..625L}.
We find temperatures of 10\,kK for \Halpha\ fibrils. This is comparable to values
found in many studies that use cloud modeling
\citep[\eg][]{1994A&A...282..939H,1997A&A...324.1183T,2011AN....332..815B} 
However, all cloud modeling requires addition of a `microturbulent
velocity' of typically 5 to 15\,km\,s$^{-1}$ as an ad hoc parameter
to derive the temperature from the observed cloud model
parameters. The profiles in our simulation have similar widths as the
observations and yield similar temperatures
without the need of invoking microturbulence.

The line formation properties of \Halpha\ show a double character. At
heights below 1\,Mm its opacity is temperature-sensitive because of
its excited lower level. At larger heights this temperature
sensitivity disappears. There the opacity is instead mainly sensitive to
density, much more akin to resonance lines, which have the ground state
as a lower level.

Our simulation supports the commonly held notion that fibrils indeed
trace the magnetic file. More precisely we show that the orientation
of the fibril-like structures in the synthetic \Halpha\ core image
align with the horizontal component of the magnetic field at the
average formation height computed for each pixel separately.
This supports the
practice of using \Halpha\ imagery to constrain and test the quality of field
extrapolations
\citep[\eg][]{2010ApJ...710.1486J,2011ApJ...739...67J,2011ApJ...742..119R}.

The dark fibrils-like structures in the simulation are
caused by ridge-like density enhancements in the chromosphere that
lead to higher formation height and hence lower emergent
intensity. It is unlikely the density ridges are the only way to create fibrils
on the Sun. The simulation has a limited resolution and spatial extent
and still lacks treatment of physical processes that might be
important in the chromosphere, such as heating due to neutral-ion drag
\citep{2010ApJ...724.1542K}.

We can now answer the question what makes \Halpha\ such a good line to
observe the magnetic structure of the chromosphere. The main reason is
that it happens to have so much opacity in the line core that it
almost always forms in the low plasma beta regime where the magnetic
field is the main structuring agent in the atmosphere. The low mass of
the hydrogen atom makes the line wide, so that the fixed-wavelength
line-core intensity is only weakly modulated by the velocity
field. Instead, the correlation of formation height and intensity and
the constant column mass at the $\tau$=1 height make the line-core
intensity a tracer of the chromospheric mass density. It is the
variation of the mass density caused by the magnetic field, waves and
shocks that gives rise to the dramatic structures seen in \Halpha.

\begin{acknowledgements}
   JL recognizes support from the Netherlands Organization for
  Scientific Research (NWO).
This research was supported by the Research Council of Norway through
 the grant ``Solar Atmospheric Modelling'' and 
 through grants of computing time from the Programme for
 Supercomputing.
The Swedish 1-m Solar Telescope is operated by the Institute for Solar
Physics of the Royal Swedish Academy of Sciences in the Spanish
Observatorio del Roque de los Muchachos of the Instituto de
Astrof\'{\i}sica de Canarias. \edt{We thank R.J.~Rutten for illuminating discussions.}
\end{acknowledgements}

\bibliographystyle{aa} 
\bibliography{%
abbett,%
avrett,%
bostanci,%
carlson,%
carlsson,%
delacruz,%
dorfi,%
golding,%
hansteen,%
kontogiannis,%
martinez-sykora,%
mihalas,%
milkey,%
neckel,%
leenaarts,%
lockyer,%
reardon,%
rutten,%
rouppe,%
rybicki,%
scharmer,%
schmieder,%
skartlien,%
uitenbroek,%
vernazza,%
wedemeyer%
}

\end{document}